# Femtosecond laser induced surface modification for prevention of bacterial adhesion on 45S5 bioactive glass

Shazia Shaikh[1], Deepti Singh[2,5], Mahesh Subramanian[2,5], Sunita Kedia[1,*], Anil Kumar Singh[3], Kulwant Singh[4], Nidhi Gupta[6], Sucharita Sinha[1]

[1] Laser & Plasma Surface Processing Section, Bhabha Atomic Research Centre, Mumbai 400085 India
[2] Bio-Organic Division, Bhabha Atomic Research Centre, Mumbai 400085 India.
[3] Laser and Plasma Technology Division, Bhabha Atomic Research Centre, Mumbai 400085 India.
[4] Materials Science Division, Bhabha Atomic Research Centre, Mumbai 400085 India.
[5] Homi Bhabha National Institute, Training School complex, Anushaktinagar, Mumbai 400094 India.
[6] Technical Physics Division, Bhabha Atomic Research Centre, Mumbai 400085 India.
[*] Corresponding authors, Email: skedia@barc.gov.in

**Abstract**

Bacterial attachment and biofilm formation on implant surface has been a major concern in hospital and industrial environment. Prevention of bacterial infections of implant surface through surface treatment could be a potential solution and hence this has become a key area of research. In the present study, the antibacterial and biocompatible properties of femtosecond laser surface treated 45S5 bioactive glass (BG) have been investigated. Adhesion and sustainability of both gram positive *S. aureus* and gram negative *P.aeruginosa* and *E. coli nosocomial* bacteria on untreated and laser treated BG samples has been explored. An imprint method has been used to visualize the growth of bacteria on the sample surface. We observed complete bacterial rejection potentially reducing risk of biofilm formation on laser treated surface. This was correlated with surface roughness, wettability and change in surface chemical composition of the samples before and after laser treatment. Biocompatibility of the laser treated BG was demonstrated by studying the anchoring and growth of human cervix cell line INT407. Our results demonstrate that, laser surface modification of BG enables enhanced bacterial rejection without affecting its biocompatibility towards growth of human cells on it. These results open a significantly potential approach towards use of laser in successfully imparting desirable characteristics to BG based bio-implants and devices.

1. Introduction:

Increasing number of premature implant failures has triggered development of techniques to reliably minimize such occurrences. Two major causes for implant failures are aseptic loosening and infections [1, 2]. Among these two, higher percentages of unsuccessful biomaterial implantation are attributed to infections [2]. Infections at implant site can be described in two stages. Stage one is the initial attachment i.e. interaction between the bacterial cell surface and implant material surface. Stage two involves specific and non-specific interactions between proteins on the bacterial surface structures and binding molecules on biomaterial surface. Microorganisms predominantly grow as communities on animate and on inanimate surfaces and once they colonize, rapid multiplication of species occurs with formation of biofilm. Growth of biofilm enhances inherent resistance of the bacterial cells and leads to inflammation and consequent implant failure through poor biocompatibility and material degradation. This has serious health implications and often requires revisions of costly implant replacement surgeries.

Attachment of bacteria on surfaces is a complicated process influenced by many factors, including the inherent bacterial property, the material surface characteristics and the environmental conditions under which growth takes place. It has been shown that the initial adhesion process is governed by physical interactions between the cell and the substrate, while biological processes involves longer time scale [3]. Since the initial interaction between bacteria and the surface of the BG plays a crucial role in adhesion of the bacteria a highly characterized surface is desirable while studying this interaction process [3, 4]. Different surface properties of biomaterials such as, morphology, surface chemistry, porosity, wettability and surface roughness determines initial bacterial adhesion and growth. To prevent formation of biofilm and infections on implants it is therefore very important to address and limit bacterial adhesion in its very early stage.

Among varieties of available biomaterials, 45S5 Hench bioactive glass (BG) (45% $SiO_2$, 24.5% $Na_2O$, 24.5% CaO and 6% $P_2O_5$) was the first biomaterial capable of bonding to bone, invented by Prof. Larry Hench at University of Florida in 1969 [5]. Since invention, BG has shown promising applications in dental and orthopedic fields. BG bonds to both soft tissue, as well as, hard tissue [5].

The bone bonding mechanism has been attributed to the formation of bone like apatite i.e. hydroxyapatite on the surface of BG in simulated body fluid. To prevent bacterial infection of BG implants it is essential to discourage initial adhesion of bacterial cells on implant surfaces. For this, several approaches including impregnation of implants with antibiotics, coating with antibacterial metals such as copper or silver [6,7], chemical treatments [8], plasma-assisted modification [9], coating with self-assembled monolayer [10], and manipulating surface topography [11], have been undertaken. Hu *et. al.* studied antibacterial property of Hench BG against three types of bacteria [12]. Bellantone *et. al.* doped BG with $Ag_2O$ to improve its antibacterial property [13]. Gorriti *et. al.* incorporated $B_2O_3$ into BG to enhance antimicrobial activity [14]. Dineva *et. al.* chemically treated BG and evaluated bacterial adhesion [8]. Raphel *et. al.* discussed possibility of inhibiting microbial cell growth on BG by multifunctional orthopedic coating [6]. Begum *et. al.* correlated antibacterial activity of BG with change in environmental pH [15]. Recently, Prabhakar *et. al.* used powdered dentin with BG and increased antibacterial efficiency of the sample [16]. Despite various techniques tried for bacterial rejection [9], complete inhibition of bacterial adherence to BG surfaces remains an unachieved goal making this an active area of research. Improvements over existing techniques are desired that can completely prohibit bacterial growth under a wide range of conditions.

In this present work, antibacterial and biocompatibility properties of femtosecond (fs) laser treated BG have been investigated. Surface topography of BG samples synthesized by melt-quench technique was modified using Titanium: Sapphire femtosecond pulsed laser by direct laser writing technique [17]. In addition to having several advantages such as, being a single step and chemical free process, ultra short pulsed laser based surface modification allows surface treatment localized both, in time and space. As reported elsewhere, femtosecond laser induced surface texturing is free from heat affected zones and modification is localized within the focal volume [18]. However, when BG surface was treated using a nanosecond pulsed laser collateral damage induced due to accumulation of heat has been reported [19]. Femtosecond laser treated BG samples have shown enhanced bio-integration in terms of superior and faster growth of crystalline hydroxyapatite layer in comparison to the sample treated using nanosecond laser beam. Here, we report our results on a number of BG samples which were surface textured using femtosecond laser. Laser treatment modified surface topography, increased roughess and hence the effective surface area of the BG sample. Attachment of three most common bacteria: *Staphylococcus aureus (S. aureus), Pseudomonas aeruginosa* (*P. aeruginosa*) and *Escherichia coli (E. coli)* has been investigate on untreated and laser treated BG samples. Interaction between the bacterial cells and BG surface primarily depends on micron and sub-micron levels of surface roughness. Hence, surface roughness parameters of the laser treated BG samples have been measured using 3D optical profilometer. In our study, samples with highest achieved average surface roughness (~ 6.52 μm) completely inhibited attachment of all three bacteria. X-ray Photoelectron Spectroscopy (XPS) results revealed the presence of calcium hydroxide on femtosecond laser modified BG surface, whereas no such signatures were found on untreated BG surface. Our X-ray Diffraction studies also confirmed formation of calcium hydroxide phase post laser treatment of BG samples. Wettability of the sample surface increased after laser treatment. Biocompability of BG post laser treatment has also been evaluated employing human cervix cell line INT407. Adhesion of INT407 on laser treated surface was equivalent to that of untreated surface of BG sample. This indicates that while bacterial attachement could be inhibited, biointegration efficiency of the BG remained intact after laser surface modification. Hence, femtosecond laser treatment can serve as an efficient and facile technique to modify surface of BG, successfully preventing bacterial adhesion. Femtosecond laser treatment can thus serve as a potential technique in generating antibacterial surface for implant devices, without compromising the bio-compatibility and bio-integration of such BG implants and devices.

## 2. Experiment:

*2.1 Materials and Methods -*

All the materials used in the study were of analytical grade or higher. Yeast extract, Tryptone and Agar were from Becton-Dickinson and Company, Churchton, MD, USA. Sodium chloride was obtained from Chemco Fine Chemicals, Mumbai, India. The animal cell culture media, trypsin and antibiotics were from Invitrogen, Carlsbad, USA. Hoechst 33342 was from Sigma Aldrich, St. Louis, USA.

*2.1.1. Bio-active glass preparation -*

Using melt-quench process 45S5 BG having nominal composition $SiO_2$ (45 wt %), $Na_2O$ (24.5 wt %), $CaO$ (24.5 wt %), and $P_2O_5$ (6 wt %) was synthesized [12]. For 100 gm glass batch, $SiO_2$, $CaCO_3$, $NH_4H_2PO_4$ and $Na_2CO_3$ were mixed and calcined at 900°C for 12 h. The calcined charge was melted in ambient at 1500°C in a Pt-Rh crucible. The fabricated glass was cut into 2 mm thick circular disks having a diameter of 1.5 cm. Flat surfaces of the sample were mechanically polished and used for laser treatment.

*2.1.2. Femtosecond laser induced surface modification -*

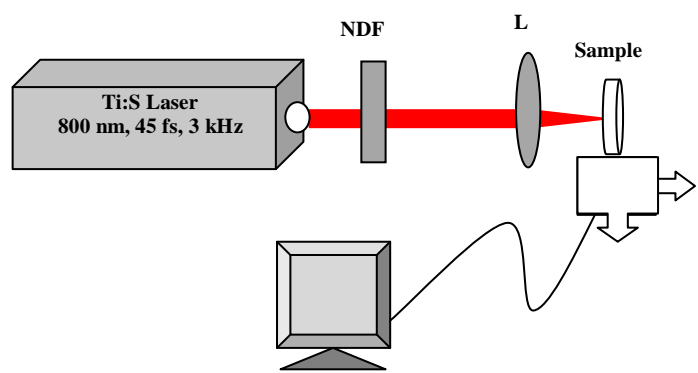

*Figure 1: Illustration of the experimental arrangement used for laser treatment of BG. NDF is neutral density filter, L is 5 cm focal length lens, and sample was fixed on a computer controlled XY-stage*

Fig. 1 illustrates the experimental arrangement used for laser surface texturing of BG. For surface modification, beam from a femtosecond Titanium: Sapphire laser delivering pulses of pulse width 45 *fs* and at repetition rate 3 kHz, at a wavelength of 800 nm was focused on to the polished surface of the BG using a 5cm focal length lens. Typical diameter of the focal spot was ~ 90 µm and value of optimized average laser power employed for surface treatment ranged between 180mW to 200mW. This corresponds to an average laser fluence of 0.94 $J/cm^2$ to 1.0 $J/cm^2$ incident on the BG sample. Surface modification or laser writing speed was controlled using a computer controlled XY-translational stage on which the BG sample was mounted. The sample was scanned in a plane perpendicular to the direction of laser beam. Sample scanning speed has been varied between 5 µm/sec to 55 µm/sec thereby generating samples with different surface roughness. Typically, an area of 5 mm x 5 mm was surface modified through femtosecond laser treatment of each BG sample. Results of untreated BG referred as S1 and samples, laser treated at scan speeds 35µm/sec (S2) and 15µm/sec (S3) are discussed here.

### 2.1.3. Surface analysis -

Surface topography of BG samples before and after laser treatment and after bacterial tests was observed under scanning electron microscope (SEM) (Carl Zeiss EVO 40 SEM).

Surface roughness parameters of untreated and laser treated BG sample were measured using a 3D-optical profilometer (Taylor-Hobson). Standard surface roughness parameters such as, arithmetic mean ($R_a$), root-mean square deviation ($R_q$) and peak to valley distance ($R_t$) also called maximum roughness were measured [4, 20].

The wettability of BG surface before and after laser treatment was studied by water contact angle measurement. The measurements were repeated at least 3 times for each sample.

XPS analysis of the BG samples before and after laser surface treatment was performed using a VG make model CLAM-2 hemispherical analyser with Al K$\alpha$ source (1486.6eV) having a step size of 0.3 eV.

X-ray diffraction (XRD) analysis of samples was carried out using PANalytical MRD system. The measurements were done using CuK$\alpha$ radiation of wavelength 1.54 A$^o$ in out of plane geometry over the 2$\theta$ range of $20^0$-$35^0$ at angle scanning speed of $2^o$/min.

### 2.2. Antibacterial test -
### 2.2.1. Preparation of medium -

Bacterial growth medium (Nutrient broth/agar) was prepared by dissolving Yeast extract (0.5%), Peptone (0.5%) and sodium chloride (0.5%) in ion free water. The pH was adjusted to 7.4. This was sterilized by autoclaving at 121 °C at 15 psi for 15 min.

*2.2.2. Bacterial cultures -*

The bacterial cultures used in the study were obtained from the repository for microorganisms, Institute of Microbial Technology, Chandigarh, India. The organisms used in this investigation are *Staphylococcus aureus* ATCC 6538P, *Pseudomonas aeruginosa* ATCC 19154, and *Escherichia coli* K Strain BW25113.

*2.2.3. Evaluation of bacterial adherence -*

Bacteria were grown overnight in Nutrient broth at 37 °C. The bacterial solution for the experiment was prepared by pelleting 1 mL overnight growth by centrifugation. The cell pellet was washed once with sterile water and re-suspended in 20 mL sterile water in a 100 mm sterile petri dish inside a laminar flow cabinet. This solution contained ~ $5 \times 10^7$ cells/mL. The BG discs that had been surface modified (area = 5 x 5 mm) using laser irradiation were wiped free of any contaminants with 70% ethanol, air dried inside the laminar hood and immersed in the bacterial solution and incubated for 2 h. After incubation the discs were carefully picked up with a sterile forcep, mildly rinsed in sterile water to remove the loosely adhered bacteria. The excess liquid was drained off. Imprint of the surface of the discs were taken on nutrient agar plates by precisely placing and pressing the discs gently once. The discs were removed with utmost care immediately. The agar plates that had the imprint of the circular discs (modified replica plate technique) were incubated for 24 h at 37 °C. The growth of the adhered bacteria was observed visually and documented by a Kodak Gel logic imaging system revealing the adherence pattern on the BG samples being tested.

*2.3. Growth of mammalian cells on BG surface -*

INT 407 cells (cervix cells of human origin) were grown in DMEM medium with antibiotic (Penicillin 100 units/mL and Streptomycin 100 μg/mL) and antimycotic (amphotericin B 0.25 μg/mL). The medium was supplemented with 10% FBS (Fetal Bovine Serum). The cells were grown at 37 °C under 5% $CO_2$. The circular 45S5 disc was placed in a 60 mm tissue culture dish. This BG disc had half its surface untreated, while remaining half had its surface modified by laser. The disc was placed in such a way that the surface to be tested for adherence of the cells was facing upward and not towards the bottom of the dish. Growth medium containing 1x $10^6$ cells was added till the disc was immersed in the medium (10 mL). The cells were allowed to attach and grow overnight in a 37 ºC incubator with 5% $CO_2$ atmosphere. Next day DNA binding dye Hoechst 33342 was added directly in to the medium (5 μM final concentration) and incubated for 10 min. The disc was lifted with a sterile forcep placed face down on a glass slide and observed under a fluorescence microscope (Carl Zeiss LSM 780) equipped with UV filters for the presence of cells.

3.  Results :

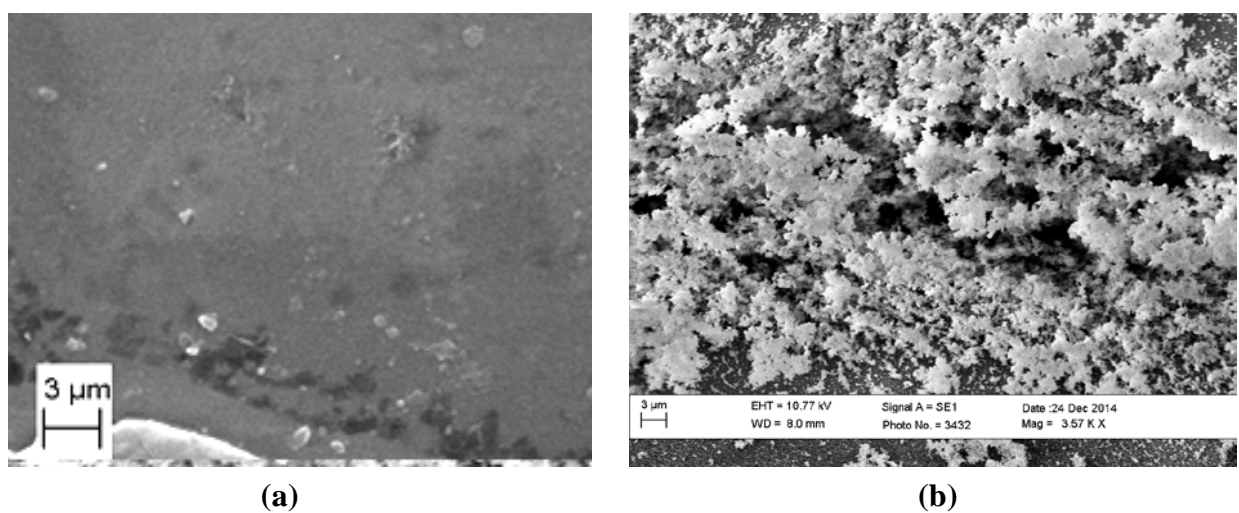

*Figure 2: SEM image of (a) untreated and (b) laser treated BG*

Figs. 2a and 2b are typical SEM images of pristine and laser treated BG, respectively. Microstructures generated on the surface of the BG after laser treatments are clearly visible in Fig. 2b. Surface roughness and hence the effective surface area of the BG samples distinctly increased after laser treatment.

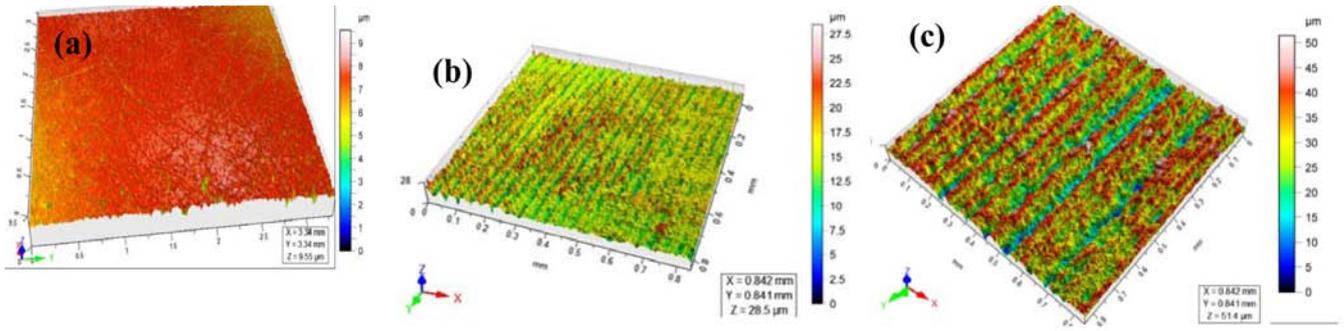

*Figure 3: Optical profilometer image of BG (a) untreated (S1), laser treated at fluence 1.0 J/cm$^2$ and sample scanning speed of (b) 35 μm/sec (S2), and (c) 15 μm/sec (S3)*

Figs. 3a, 3b and 3c are surface scanned optical images of sample S1, S2 and S3, respectively. Color variation in the images indicates modified topography of the BG after laser treatment. Roughness parameters of the untreated and laser treated BG are summarized in table-I.

| Sample | $R_a$ (μm) | $R_q$ (μm) | $R_t$ (μm) |
|---|---|---|---|
| S1 | 0.42 | 0.32 | 7.50 |
| S2 | 1.88 | 2.54 | 23.5 |
| S3 | 6.25 | 7.74 | 43.7 |

*Table-I: Roughness parameters of different BG samples*

Roughness of the sample enhanced significantly when BG was translated at lowest scanned speed, referred to as sample S3. Slower movement allowed longer interaction time to the laser pulses with the BG. Multiple laser pulses deposited their energy on BG surface and when temperature exceeded melting point of the sample, melting occurred on BG surface. Associated hydrodynamic instability in the melt volume resulted in modification of surface [21]. Subsequent interaction of the laser pulses with the roughened surface of the BG caused non-uniform energy deposition. Higher laser energy was deposited in the valley regions in comparison to peaks formed on the surface. This non-uniform energy distribution resulted in a temperature gradient and hence variation in surface tension on BG surface, temperature of the hills being lower than that of the valleys. As, energy deposition on the sample increased, melted BG flowed from low surface tension region (valley region) to high surface tension region (hill) and further elongated the microstructures, as evident in case of S3. Repetition rate of used femtosecond laser was 3 kHz and focal spot size of the laser beam on BG surface was 90 μm. Samples S2 and S3 were scanned at speed of 35 μm/sec and 15 μm/sec, respectively. Hence, on an average S2 received lesser (~ 7,500 pulses/spot) laser pulses in comparison to S3 (~ 18,000 pulses/spot). Therefore, generation of larger temperature gradient on S3 surface can be expected. This caused formation of high hills with deep valleys on surface of S3. In Fig. 3c, $R_t$ for S3 was measured as 43.7 μm which was significantly larger than $R_t$ for S1 (7.50 μm) and S2 (23.5 μm).

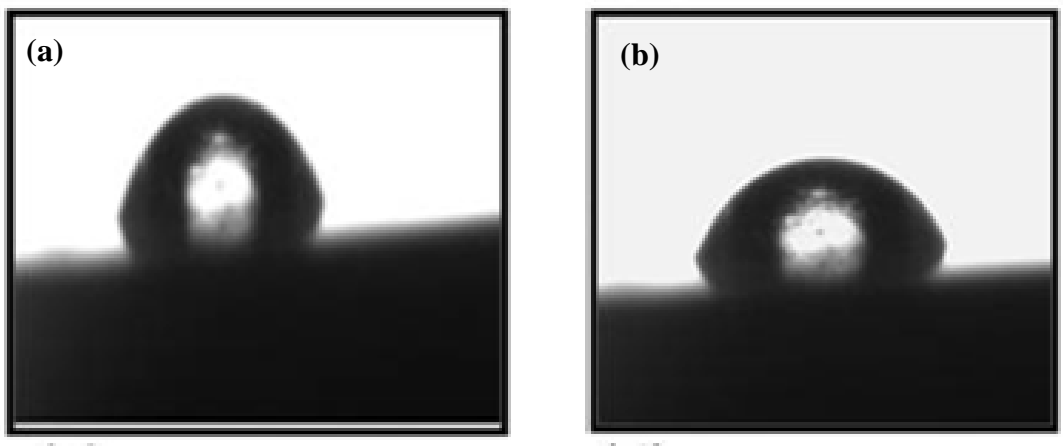

*Figure 4: Photographs of water contact angle measured for BG (a) S1 (73°), and (b) S3 (43°)*

Figs. 4a and 4b are the photographs of water contact angle measurement of S1 and S3, respectively. The water contact angle of pristine BG was 73° indicating hydrophilic nature of the surface. The contact angle decreased to 34° for S3. Therefore, wettability of the BG samples was found to increase on laser treatment.

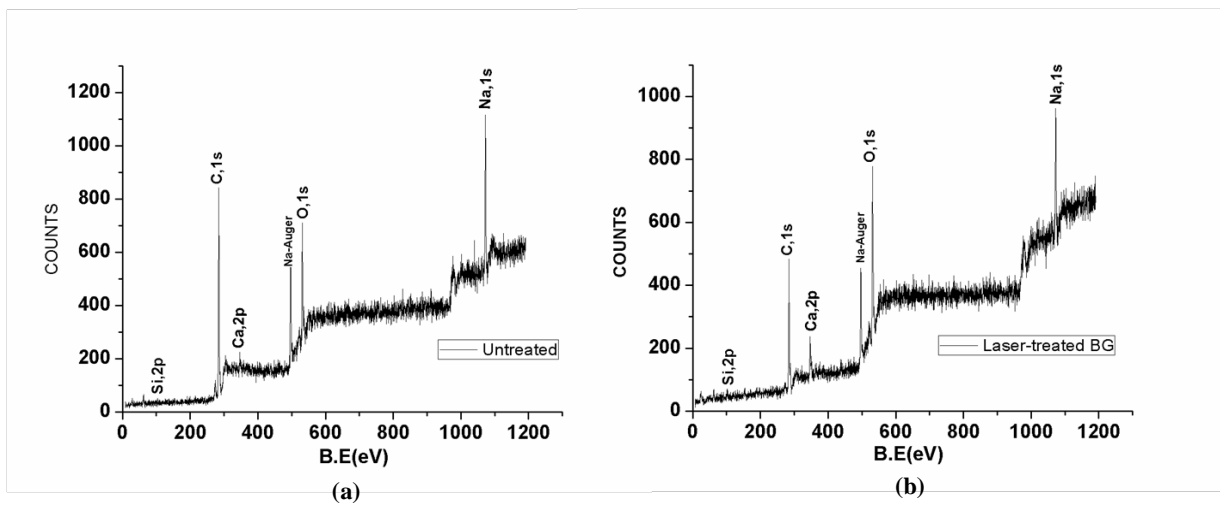

*Figure 5: XPS survey spectra of (a) untreated and (b) fs-laser treated BG (S3)*

Surface compositions of BG before and after fs-laser treatment were analyzed by X-ray Photoelectron Spectroscopic technique. The XPS survey spectra of pristine BG and laser modified BG are shown in fig.5a and 5b, respectively. Peaks for the possible constituent elements have been identified and marked on these scans in fig.5. A major difference observed between survey spectra taken before and after laser treatment were the prominent peaks of Ca and O for the laser treated BG representing increased presence of calcium and oxygen post laser treatment. Also, less intense peak of carbon was observed on the surface of laser modified BG in comparison to untreated BG. Lower carbon presence on laser treated BG surface could be due to surface sputtering and cleaning on account of laser irradiation. Carbon contamination of pristine BG surface could also be responsible for absence of detectable peak of Si in case of untreated BG samples. Spectral peaks around 1072.5eV, 531.5eV, 346.5eV, 284.4eV and 102.1eV were assigned to the Na1s, O1s, Ca2p, C1s, and Si2p photoelectrons, respectively. The peak at 469.5eV is assigned to Na KLL Auger transition. In order to have a better understanding of the effect of laser treatment on BG O1s, C1s and Ca2p spectra were deconvoluted into their components.

The O1s peak provides very useful information about the oxides in glasses. Silica based glasses generally consist of two kinds of bonding with oxygen atoms i.e. bridging oxygen bond and non-bridging oxygen bond due to presence of alkali network modifiers (Na and Ca) in glass composition. The high resolution O1s spectra for both untreated and fs-laser treated BG have been deconvoluted into four components, as shown in fig.6. (a). For fs-laser treated and untreated BG non bridging oxygen peak appears at 528.6eV and 529.5eV, respectively [22]. Bridging oxygen peak was observed as expected at 531eV for untreated BG [22]. Na-Auger photoelectron peak was observed for both laser treated and untreated BG at binding energy values 535.7eV and 535.9eV [23]. Peak at 533.3eV observed in the case of fs-laser treated sample can be attributed to the presence of OH group [24]. Peak at 531.6eV and 531.8eV, is an indication of the presence of C=O, hydrocarbonaceous contamination on untreated and laser treated BG respectively [25].

In Fig.6b are shown the deconvoluted spectra of Ca2p for untreated and fs-laser treated BG samples. For untreated BG the Ca2p spectra was deconvoluted into two components having peak positions 346.8eV and 351eV, corresponding to 2p states of CaO and $CaCO_3$. Spectra for *fs*-laser treated BG when deconvoluted gave three components with peak positions at 346.5eV, 347.4eV, and 350.5eV. Peak appearing around 346.5eV can be attributed to $2p_{1/2}$ of $Ca(OH)_2$ or $2p_{3/2}$ of CaO, peak at 347.4eV can be associated with $2p_{3/2}$ of CaO or $CaCO_3$, while peak around 350.5eV can be assigned to $2p_{1/2}$ of CaO or $Ca(OH)_2$ [26]. Hence, these three peaks can be attributed to either $2p_{1/2}$ or $2p_{3/2}$ states of Ca corresponding to presence of $Ca(OH)_2$, CaO or $CaCO_3$.

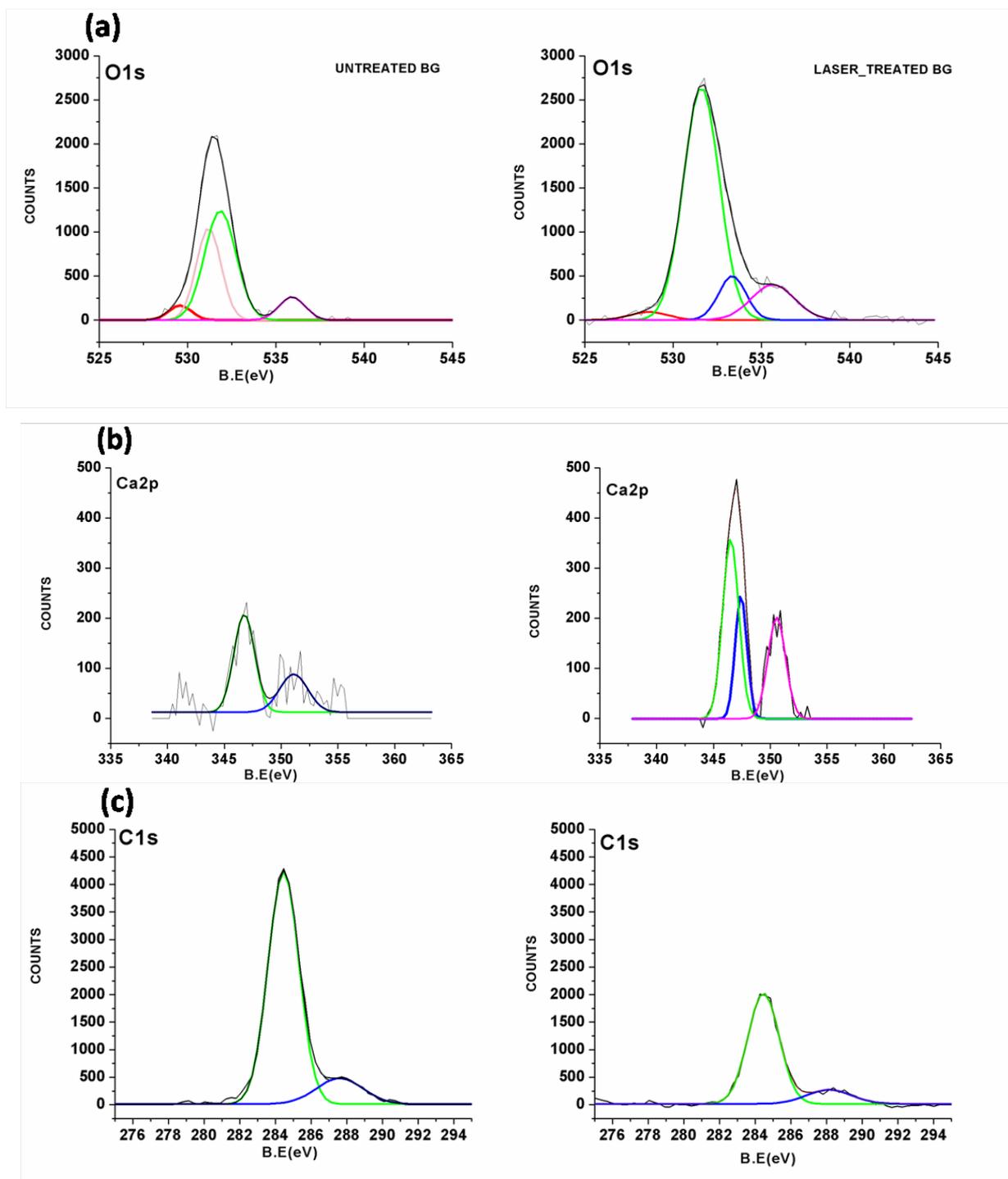

*Figure 6: High resolution XPS spectra of (a) O1s, (b) Ca2p, (c) C1s, of BG untreated surface (left) and fs-laser treated surface (right)*

The high resolution XPS spectra for C1s for both pristine and laser modified BG are shown in fig.6. (c). The carbon spectral components observed at binding energy values 284.8eV and 287.7eV, correspond to C-C and C=O, respectively [24].

Si 2p spectra for laser modified BG showed two peaks at 102.1eV and 104.8eV corresponding to non bridging oxygen and bridging oxygen atoms when bonded to Si atom [22]. Surface contamination of untreated BG samples could be the likely reason for Si 2p spectra not being detected in XPS of pristine BG samples i.e. prior to laser treatment.

Dominant peak for Na1s for both untreated and fs-laser treated BG at 1072.5eV and 1072.8eV, respectively indicated the presence of $Na_2O$. While, binding energy value for elemental Na1s is 1071.5eV, 1eV shift of binding energy indicates oxide formation [27].

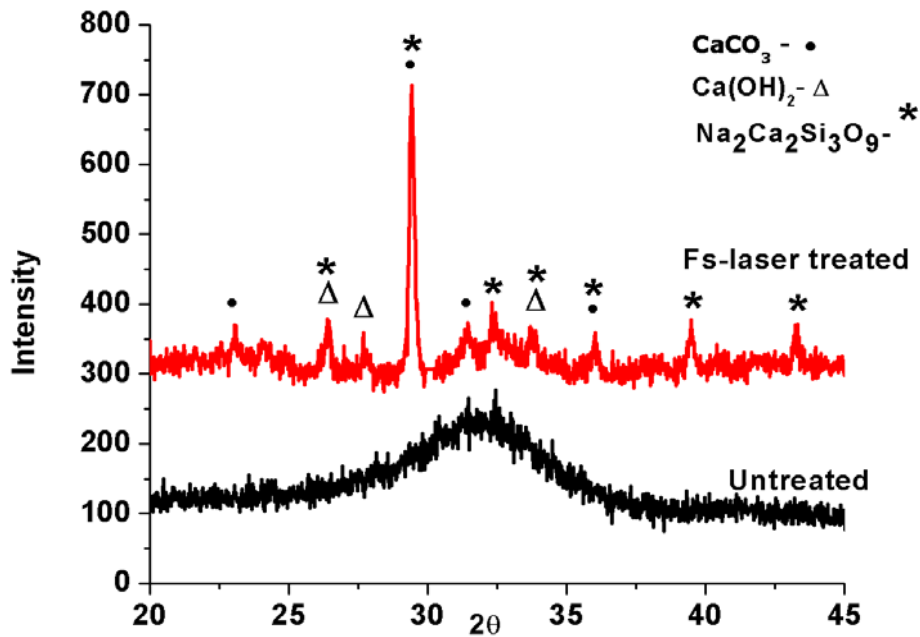

*Figure 7: X-ray diffraction pattern of untreated and fs-laser treated BG*

Fig.7 is shown XRD pattern of untreated and fs-laser treated BG, respectively. The broad peak centered at 32° for untreated BG confirms the amorphous nature of BG (black) whereas sharp peaks observed in XRD pattern on laser treatment (red), suggest surface crystallization of BG post fs-laser irradiation. Most of these sharp peaks obtained after laser treatment matched with JCPDS # 22-1455, 85-1108 and 87-0674 confirming $Na_2Ca_2Si_3O_9$, $CaCO_3$ and $Ca(OH)_2$ crystalline phase formation, respectively. As demonstrated by various studies, BG is prone to crystallization on heat treatment and forms sodium-calcium silicate phase [28]. Heat treatment at high temperature generally shows the presence of $Na_2Ca_2Si_3O_9$ phase, a prominent crystalline phase of 45S5 bioglass. In our process of laser treatment of BG using a femtosecond laser in addition to sodium-calcium silicate phase we have also observed presence of $CaCO_3$ and $Ca(OH)_2$ crystalline phases. Chemical reaction of BG surface with atmospheric environment during laser surface treatment is expected to have resulted in formation of $CaCO_3$ and $Ca(OH)_2$. Our XRD data thus confirms presence of both $CaCO_3$ and $Ca(OH)_2$ on laser treated surface of BG as was also indicated by our XPS investigations.

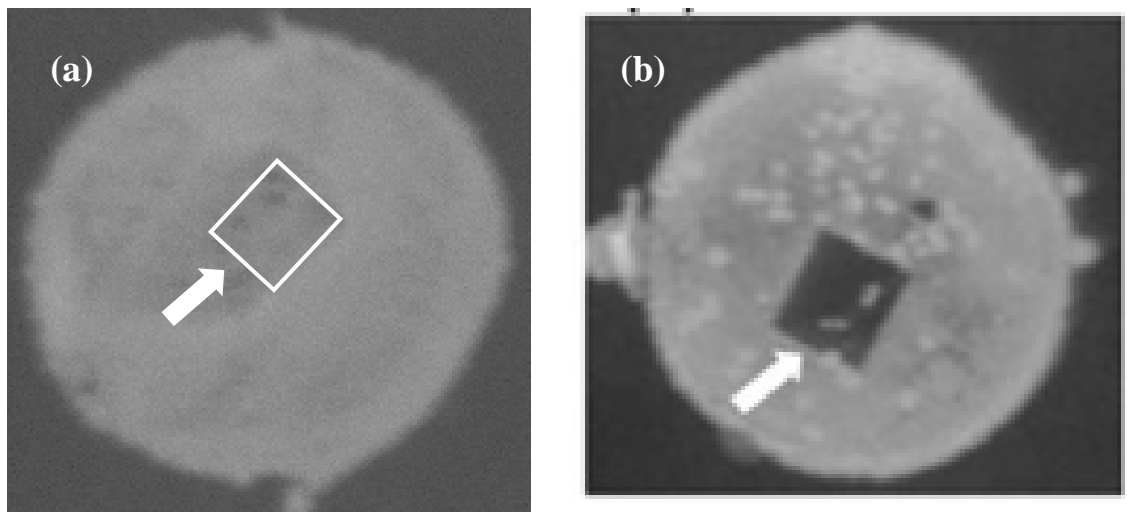

*Figure 8: Photograph of Nutrient agar plate showing replica imprint of BG with laser treated zone (a) S2 and (b) S3 after allowing S. aureus to incubate for 24 hrs.*

Precise imaging of bacterial adhesion and growth on an implant surface is essential for determining the success and effectiveness of the implant material. We have employed a novel way to test the growth of bacteria adhered to the surface of the implant material based on taking an imprint on the solidified growth medium. For this, BG sample with laser treated zone was immersed in *S. aureus* bacterial solution for 2 hr and imprint of the sample was taken on nutrient agar plate which was incubated for 24 hrs. Figs. 8a and 8b are the photographs of nutrient agar plate showing replica imprint of S2 and S3, respectively. The laser treated region is marked with arrow in both the figures. Adhesion and agglomeration of *S. aureus* was observed on untreated and laser treated regions in S2, as seen in Fig. 8a. Unlike this, adhesion of *S. aureus* significantly reduced on S3, as observed by clear zone in Fig. 8b. Only 10% of total laser treated area allowed colonization of *S. aureus* on S3 surface as evident in the figure.

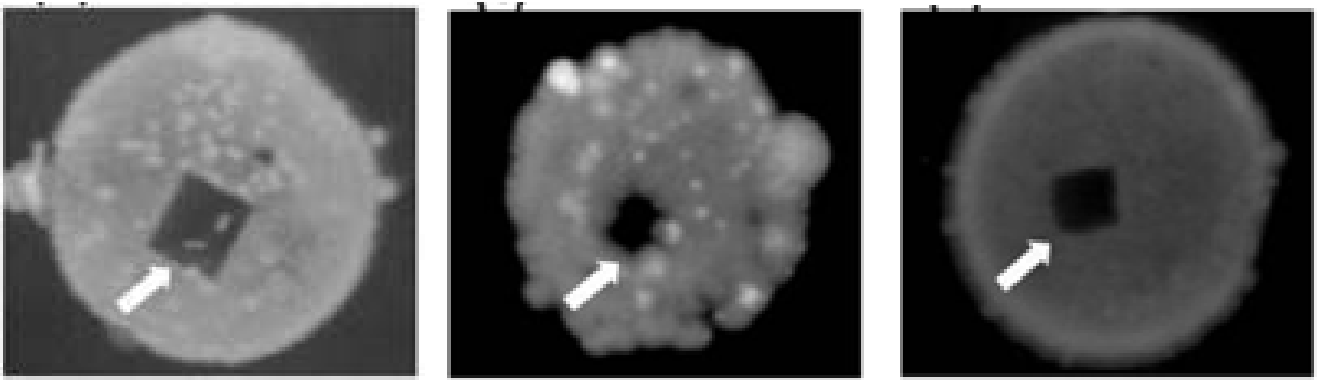

*Figure 9: Photograph of Nutrient agar plate showing replica imprint of S3 adherence of (a) S. aureus , (b) P. aeruginosa, and (c) E. coli.*

Figs. 9a, 9b and 9c are the photographs of nutrient agar plate showing replica imprint of sample S3 incubated with S. aureus, *P. aeruginosa*, and *E. coli*, respectively. The laser treated portion of BG successfully obstructs adhesion of *P. aeruginosa*, and *E. coli* as observed by a clear zone (marked by arrows), in Fig. 9b and 9c, respectively. However, all three bacteria adhere and cohere to the untreated region of the BG.

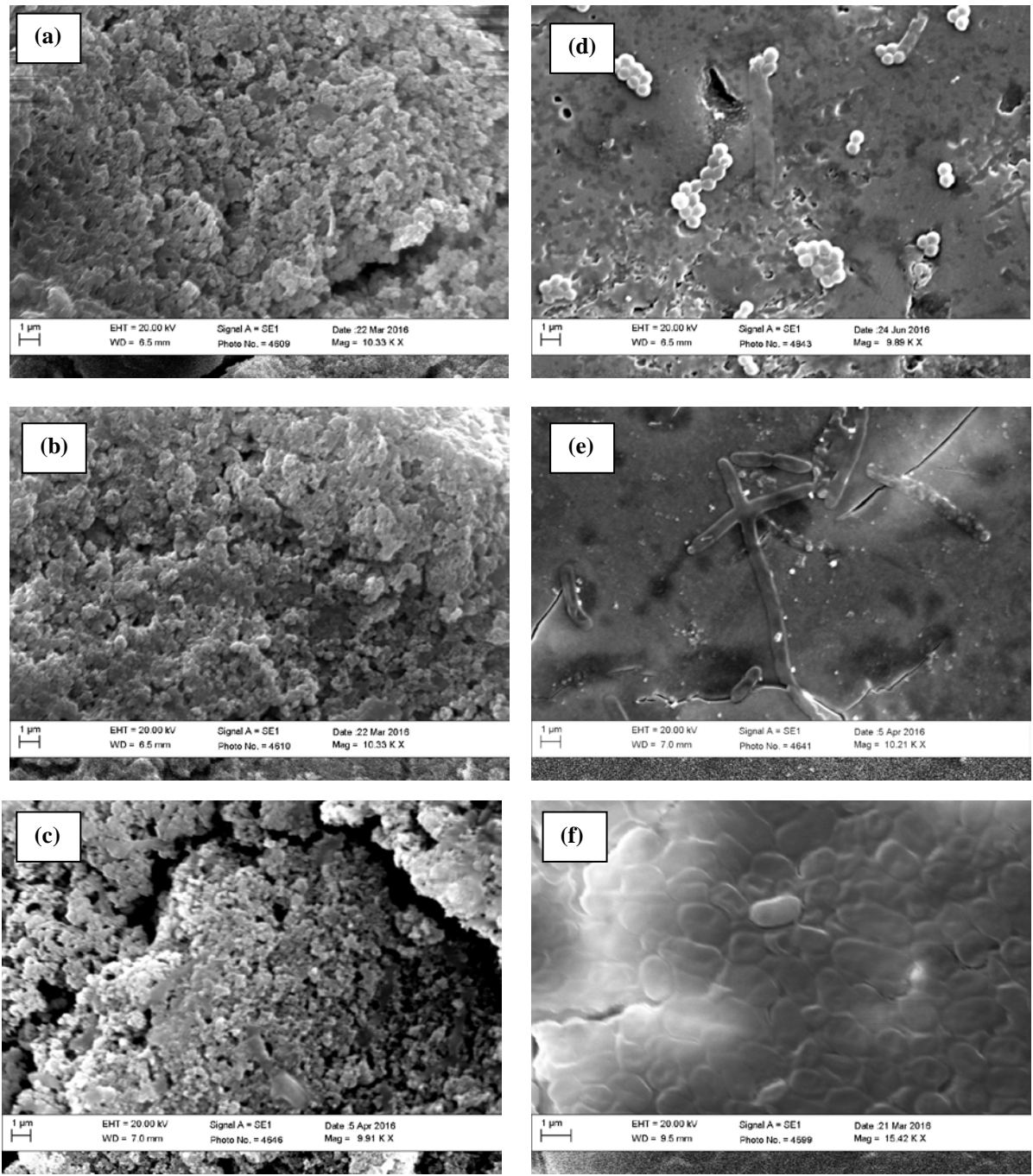

*Figure 10: SEM images of laser treated area of S3 showing no adherence of (a) S. aureus , (b) P. aeruginosa, and (c) E. coli, and SEM images of untreated BG showing adherence of (d) S. aureus , (e) P. aeruginosa, and (f) E. coli*

The antibacterial property of the laser treated BG was also confirmed by independent investigation employing sample imaging under scanning electron microscope (SEM). The bacteria were allowed to adhere to the BG discs for 2 h after which the discs were rinsed mildly in sterile water and air dried for 10 min. The surfaces of the discs were gold coated by sputtering and bacterial adhesion was observed under SEM. Physical adherence of the different bacteria to the laser treated vis-a-vis untreated surface was investigated. As seen in the fig. 10, in all the 3 cases, no bacterial cells were seen on the laser treated surface (figures 10a, 10b, and 10c) while bacterial adhesion was clearly observed on the untreated surface (figures 10d, 10e, and 10f).

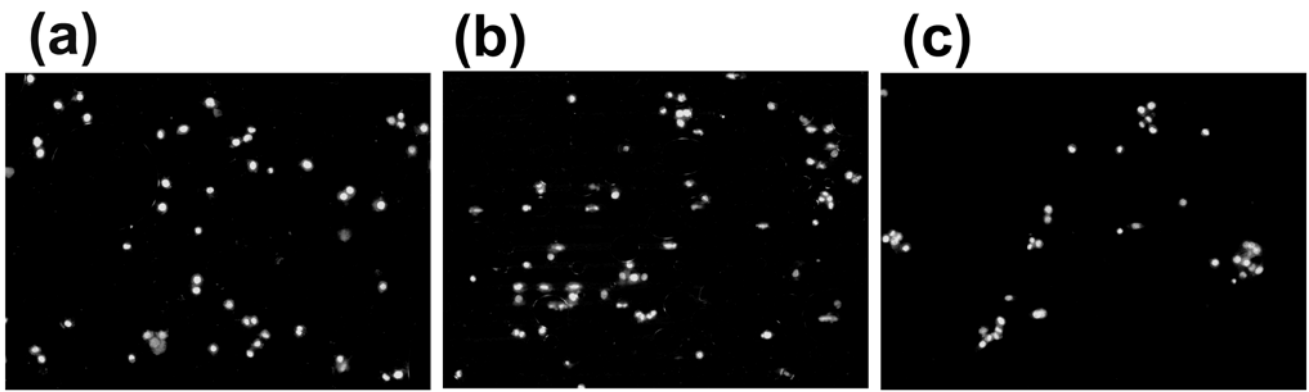

Figure 11. INT407 Cells growing on (a) untreated, (b) laser treated BG surfaces (S3) and (c) animal cell culture Petri dish

Successful bio-integration of implants depends on their bio-compatibility. Hence, bio-integration is a property that needs to remain intact post laser surface treatment. Therefore, adhesion and growth of a typical human cell type INT407 was tested on the laser treated BG surface and compared with an untreated surface. It was evident that there existed no difference in the adherence and growth of INT407 cells on the S1 (Fig. 11a) and S3 (Fig. 11b) surfaces. Adherence of human cell is a desirable property as under real clinical condition when the host cells and tissues accept the laser treated implant surface as a scaffold to adhere and grow, bio-integration of implant is successful. We also confirmed that the growth observed on the 45S5 surface was comparable to the growth of the same cells on animal tissue culture dish that has poly-lysine coating which provides an anchoring surface for the cells to grow as shown in Fig. 11c.

4. **Discussions :**

Topography mediated bacterial adhesion has been studied in the light of a comprehensive description of the surface architecture of 45S5 Hench BG substrate. The effect of surface roughness on adherence of *S. aureus, P. aeruginosa* and *E. coli* bacteria has been evaluated. *S. aureus* is a gram positive bacterium spherical in shape with typical size ~ 0.5-1.0 μm. This bacterium accounts for ~70% of implant infections [29]. *P. aeruginosa* is a gram negative rod shaped bacterium with size around 2.0 μm. These are notorious for causing nosocomial infections (~ 8 %) that are pertinent to infections associated with implant devices. *E. coli* is a rod shaped, gram negative bacterium causing gastroenteritis and it is one of the most established model organism for bacterial research. In our present study, *S. aureus* adhered and colonolized on the BG surfaces with $R_a$ 0.42 μm (S1) and 1.88 μm (S2). However, for the sample (S3) with $R_a$ 6.25 μm, *S. aureus* nucleated only on 10% of total laser treated area, as seen in Fig. 8b. Additionally, S3 rejected attachment and bonding of *P. aeruginosa* and *E. coli* bacteria completely, as observed in Fig. 9b and 9c, respectively. Maximum roughness ($R_t$) of S3 was measured as 43.7 μm which was much larger than $R_t$ for S1 (7.50 μm) and S2 (23.5 μm). Red color depicted in Fig. 3b and 3c indicates location of peaks on the surface of the laser treated samples. Summit density; that is a measure of the number of peaks per unit area is expected to be much higher on the surface of S3 in comparison to surface of S2 as confirmed from the distribution of red color in Fig. 3c and 3b, respectively. Increased surface roughness for S3 resulted in discontinuous contacts, and thus reduced the bonding opportunity of bacteria [30]. Additionally, comparatively dense features of high peaks associated with deep valley restricted bacteria to reach the BG surface and limited their agglomeration on S3 surface. In comparison, maximum roughness in case of S2 was relatively lower and summit density also significantly less. Hence, for this sample bacteria could adhere on the BG surface and could agglomerate easily.

Kathryn *et. al.* reported retention of *S. aureus* bacteria on biomaterial with $R_a$ ~ 2 μm [31]. Similarly, in present case $R_a$ for S2 was comparable and it supported attachment of S. aureus. Taylor *et al* reported increase in *P. aeruginosa* adhesion on the biomaterial surface with average roughness ~ 0.04 to 1.24 μm, although in the same report it was observed that bacterial adhesion decreased notably when $R_a$ was ~1.86 to 7.89 μm [30]. This is in agreement with our observation in the current investigation where S3 sample with $R_a$ ~ 6.25 μm inhibited bacterial adhesion completely. Bagherifard *et. al.* reported adhesion and number of colony formation of *S. aureus* decreases as surface roughness increases [20]. Also, that report clearly mentions that, density of *S. aureus* on the biomaterial surface is inversely proportional to $R_q$. The roughness parameters for the sample which stunted adhesion of *S. aureus* in that work were measured as $R_a \sim 8.14$ μm, $R_q \sim 10.10$ μm and $R_t \sim 63.43$ μm. In our present work, measured roughness parameters for S3 were quite similar and the sample S3 rejected adhesion of all three bacteria completely.

In addition to surface topography antibacterial behavior of BG has also been correlated to a high pH value environment. Probable mechanisms responsible for death of bacterial cells when exposed to a strong base have also been proposed. Calcium hydroxide is a strong alkaline substance having a pH of 12.5. In an aqueous environment it dissociates into calcium and hydroxyl ions. Hydroxyl ions show extreme reactivity and have been reported to have lethal effects on bacterial cells [32]. Reported studies on antibacterial behavior of BGs have correlated it to high aqueous pH values, as an alkaline environment has been shown to be detrimental for most of the bacteria [15].

Our XPS results reveal that after fs-laser surface treatment calcium hydroxide formed on the surface of BG. whereas no signature of calcium hydroxide was detected on untreated BG surface. Our XRD results also confirmed presence of calcium hydroxide post laser surface treatment of BG surface. Calcium hydroxide has been extensively used by endodontics because of its antibacterial properties [33]. Therefore, antibacterial effect of fs-laser modified BG sample surface as observed in our study could have also been enhanced owing to the presence of calcium hydroxide on BG surface as a consequence of laser treatment. Our results suggest that surface roughness, as well as, presence of calcium hydroxide on BG surface post fs-laser treatment could collectively be responsible for the observed antibacterial behavior. Thus, both surface topography and chemical composition play a crucial role resulting in antibacterial effects of BG surfaces.

Higher wettability with adequate antibacterial property is desirable for an ideal biomaterial, as better wetness helps in tissue attachment and integration [34, 35]. As per our observation, untreated BG surface was hydrophilic in nature and wettability of the sample improved post laser treatment. While, some reports have correlated high bacterial rejection with surfaces having high hydrophobicity [36], recently Prabu *et. al.* reported bacterial rejection also by hydrophilic surface [35].

## 5. Conclusion :

Antibacterial effect of rough surface and its potential ability to reduce the risk of biofilm formation is of utmost importance. Our results demonstrated that modification of surface texture along with chemical composition holds great potential in controlling microbial attachment. Enhanced antibacterial property with intact biocompatibility of femtosecond laser treated BG, was observed. Surface roughness and wettability of the BG increased after femtosecond laser treatment. Both XPS and XRD results have confirmed presence of calcium hydroxide on the fs-laser modified BG surface. Both surface roughness and presence of calcium hydroxide on laser treated BG surface enhanced its antibacterial behavior. Complete rejection of three types of nosocomial bacteria was seen on most rough BG surface with $R_a \sim 6.54$ μm and $R_t \sim 43.7$ μm. Our observations also confirmed that attachment and growth of INT407 human cells on laser treated BG samples remained unaltered when compared to pristine untreated samples. Adherence and growth of human cells on such BG determines its biocompability enabling rapid bio- integration of these implants. Thus, surface modification of BG by femtosecond laser is an attractive means of mitigating bacterial attachment while retaining biocompatibility as revealed by the growth of INT 407 human cells.